\documentclass[fleqn,10pt]{wlscirep}
\usepackage[utf8]{inputenc}
\usepackage[T1]{fontenc}

\usepackage{array}
\usepackage{bm}

\usepackage{upgreek}
\newcommand{\tu}[1]{\textup{#1}}
\newcommand{\dvt}[1]{\frac{\mathrm{d} {#1}}{\mathrm{d} t}}
\newcommand{\dvtt}[1]{\frac{\mathrm{d}^2{#1}}{\mathrm{d}t^2}}
\newcommand{\SI}[1]{\textit{Supplementary section {#1}}}

\title{Unified Framework for Laser-induced Transient Bubble Dynamics within Microchannels}
\author[1]{Nagaraj Nagalingam}
\author[1]{Vikram Korede}
\author[1]{Daniel Irimia}
\author[1]{Jerry Westerweel}
\author[1]{Johan T. Padding}
\author[1]{Remco Hartkamp}
\author[1,*]{Hüseyin Burak Eral}
\affil[1]{Process \& Energy Department, Delft University of Technology, Leeghwaterstraat 39, 2628 CB Delft, Netherlands}
\affil[*]{h.b.eral@tudelft.nl}

\begin{abstract}
Oscillatory flow in confined spaces is central to understanding physiological flows and rational design of synthetic periodic-actuation based micromachines. Using theory and experiments on oscillating flows generated through a laser-induced cavitation bubble, we associate the dynamic bubble size (fluid velocity) and bubble lifetime to the laser energy supplied - a control parameter in experiments. Employing different channel cross-section shapes, sizes and lengths, we demonstrate the characteristic scales for velocity, time and energy to depend solely on the channel geometry. Contrary to the generally assumed absence of instability in low Reynolds number flows ($<1000$), we report a momentary flow distortion that originates due to the boundary layer separation near channel walls during flow deceleration. The emergence of distorted laminar states is characterized using two stages. First the conditions for the onset of instabilities is analyzed using the Reynolds number and Womersley number for oscillating flows. Second the growth and the ability of an instability to prevail is analyzed using the convective time scale of the flow. Our findings inform rational design of microsystems leveraging pulsatile flows via cavitation-powered microactuation. 
\end{abstract}

\begin{document}

\flushbottom
\maketitle

\thispagestyle{empty}

\section*{Introduction}
Micromachines with few mechanical components have revolutionized the areas of microelectromechanical systems (MEMS) \cite{10.1080/108939598199991,10.1002/smll.201904032,10.1089/soro.2019.0169}. The potential of bubble-powered micromachines was first realized in ink-jet printing \cite{allen1985hewlett}, later also finding other applications that require precise flow control and rapid actuation \cite{10.1039/D2LC00169A,10.1038/s41598-022-24746-w,10.1103/PhysRevApplied.6.024003}. Oscillatory flows are not limited to synthetic devices, but also invariably exist in nature, e.g., in cardiovascular and respiratory flows \cite{10.1113/jphysiol.1955.sp005276}. Yet characterization and optimization of pulsated oscillatory flow is underexplored compared to steady flows \cite{10.1002/smll.201904032}. Therefore, a unified understanding will inform both synthetic and physiological systems, encompassing particle manipulation \cite{10.1073/pnas.0811484106,10.1038/s41598-017-03114-z}, rheology \cite{10.1039/C0LC00182A,10.1038/s41598-020-68621-y}, emulsification \cite{10.1088/0034-4885/75/1/016601}, cell lysis \cite{10.1039/B715708H,10.1038/nphys148}, prilling \cite{10.1103/PhysRevFluids.6.103903}, needleless injection \cite{10.1007/s10439-019-02383-1,10.1038/s41598-020-61924-0} and gas embolotherapy \cite{10.1115/1.1824131}.

Laser-induced cavitation allows localization of high temperature and large flow velocities due to the growth and collapse of short-lived vapor bubbles. Thus, laser light can be implemented to induce flow in processes with only need for optical access \cite{10.1007/978-981-287-470-2_6-1}. For confined geometries, the arrested directions of flow and increase in bubble lifetime due to strong confinement-induced momentum dissipation allows for a simplified theoretical and experimental approach \cite{10.1063/1.870381,10.1017/S0022112009007381}. Yuan \textit{et al.} \cite{10.1016/S0017-9310(99)00027-7} laid the theoretical foundation for the dynamics of a vapor bubble in a narrow channel with a circular cross-section. They reported the internal vapor pressure of the bubble to be insignificant after the initial 10\% of the bubble's lifetime. Thus, the dynamics of the bubble is largely governed by the wall resistance of the microchannel, ambient pressure, and inertia of the liquid shortly after bubble formation. Using experiments and a numerical model, Sun \textit{et al.} \cite{10.1017/S0022112009007381} showed the role of absorbed laser energy in bubble evolution within circular microchannels with internal diameters of 25 and 50 $\upmu$m.

The key to rational control of thermocavitation driven flows hinges on understanding the dynamics of the transient bubble, the bubble lifetime, and its dependence on the laser energy supplied. The findings could potentially be applied to any microsystem that employs a periodic actuation source to transfer momentum to the fluid much faster than the timescale of flow oscillation. Given application-specific requirements, e.g., flow rates, droplet production rate and mixing \cite{10.1038/srep09942}, valve actuation time \cite{10.1039/C0LC00520G}, etc., there is a need for a unified framework with which one could \textit{a priori} design the channel geometry (cross-section shape, size and length), provided that the fluid properties such as viscosity, density and light absorbance are known. Beyond the design of \textit{lab-on-a-chip} devices, characterizing the flow and instabilities will help in the study and engineering of physiological systems, such as flows in arteries and bronchioles, by additionally incorporating the effects of channel wall compliance \cite{10.1007/s00348-003-0776-9}. Furthermore, the fundamental understanding developed can be utilized in laser-induced crystallization via thermocavitation \cite{10.1103/PhysRevLett.131.124001,10.1021/acs.cgd.2c01526,10.1021/acs.cgd.9b00362}, as the bubble dynamics will dictate hydrodynamic flows around nucleated crystals.

In this paper, combining high-speed microscopy and an analytical approach, we demonstrate a universal dependence of the magnitude and duration of the induced flow on the laser energy supplied in microchannels with circular, square, and rectangular cross-sections. Prior works discussed undistorted laminar flows \cite{10.1017/S0022112009007381,10.1088/0960-1317/15/3/028,10.1063/1.4769755,10.1063/1.5142739}, which is a convenient assumption for low Reynolds numbers ($Re<1000$) \cite{10.1146/annurev-fluid-120720-025957}. Nevertheless, we observe the occurrence of a transient flow instability even at $Re<1000$, and delineate it by characterizing two dimensionless numbers: (i) the Reynolds number ($Re$), contingent on the peak mean flow velocity and channel hydraulic diameter, and (ii) the Womersley number (\textit{Wo}), contingent on the bubble lifetime and channel hydraulic diameter. This momentary instability emerges due to the oscillatory nature of the flow \cite{10.1299/jsme1958.25.365,10.1073/pnas.1913716117}, and results in momentary unsteady velocity profiles due to the disruption of the momentum boundary layer near the channel walls \cite{10.1115/1.1490375}.

This paper first details the experiments for quantifying the dynamic bubble size and lifetime as the function of its maximum bubble size. Then the associated theoretical framework is presented to explain underlying physics. Following which, a general empirical correlation between maximum bubble size and laser energy supplied, and the minimum threshold laser energy for bubble formation is discussed. Finally, using the established theoretical framework and experiments, we characterize the flow transition limits and nature of the distorted laminar flow.\\

\section*{Methods}
\begin{figure}[!t]
\centering
\includegraphics[width=\linewidth]{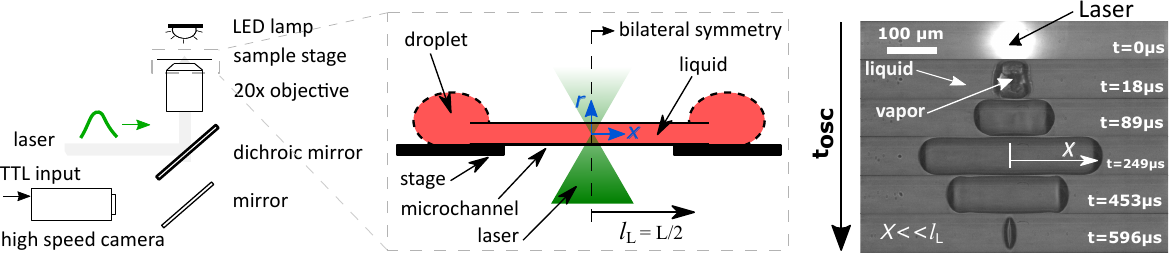}
\caption{A sketch of the experimental setup with representative images of bubble dynamics from inception to collapse - one oscillation cycle. $X$ represents the bubble size and $t_\tu{osc}$ is the bubble lifetime.}
\label{figure 1}
\end{figure}

\indent \textbf{Laser setup:} Frequency-doubled Nd:YAG pulsed laser with 532 nm wavelength, 4\,mm beam diameter and 4\,ns pulse duration. A $20\times$ objective (numerical aperture = 0.5) is used to focus the laser within the microchannel and simultaneously used for imaging \cite{10.1016/j.ohx.2023.e00415}. The images are recorded at 112,500\,frames per second using a high-speed camera. \textbf{Working fluid:} An aqueous solution of red dye (RD81, Sigma-Aldrich) with 0.5 wt\% is used to have higher absorbance to light at 532 nm. The absorption coefficient ($\alpha$) was measured to be 173 cm$^{-1}$ using a spectrometer (DR6000, Hach). The liquid was not pre-treated for dissolved gases. \textbf{Channel geometry:} The hydraulic diameter ($d_\tu{h}$) of the channels range from 66.7\,$\upmu$m to 300\,$\upmu$m, with the maximum bubble half-size, $X_\tu{max} \in [d_\tu{h}/2,2\,d_\tu{h}]$ (see Fig.\,\ref{figure 1}). Two different capillary lengths, $L=25\,\,\tu{and}\,\,50 \,\tu{mm}$, were used for all cases with the laser always focused at the geometric center. The channel specifications used in this work are summarized in Table\,\ref{channel geometry}.
\begin{table}[!h]
\centering
\caption{Microchannel specifications used in this work. Material: Borosilicate; Tolerance +/- 10\%; Manufacturer: CM Scientific.}
\begin{tabular}{>{\centering\arraybackslash}p{9em}>{\centering\arraybackslash}p{9em} >{\centering\arraybackslash}p{9em} >{\centering\arraybackslash}p{9em}} \hline
Cross-section & hydraulic diameter & Inner diameter & Wall thickness \\
 & ($\upmu$m) & ($\upmu$m)  & ($\upmu$m) \\
\midrule
Circular & 100 & 100 & 35 \\
Circular & 200 & 200 & 65 \\
Circular & 300 & 300 & 50 \\
Square & 100 & 100 & 50 \\
Square & 200 & 200 & 100 \\
Square & 300 & 300 & 150 \\
Rectangle & 66.7 & 50\texttimes100 & 50 \\
Rectangle & 126 & 80\texttimes300 & 56 \\
\bottomrule
\end{tabular}
\label{channel geometry}
\end{table}

\section*{Results and Discussion}
\subsection*{Dynamic bubble size} 
\indent Since the midpoint of the capillary and the bubble coincides, in our theoretical model we analyze only half the geometric domain owing to the bilateral symmetry, see Fig.\,\ref{figure 1}. Thus, the length of the liquid column under analysis ($l_\tu{L}$) is half the total length of the channel used, and the half-size of the bubble ($X$) is determined by the position of the vapor-liquid interface from the center as illustrated in Fig.\,\ref{figure 1}. In the post-processing of the images the parameter $X$ was estimated by dividing the bubble volume with the area of the channel cross-section (see \SI{1} \cite{SI}). This approach of determining $X$ ruled out the effects of bubble end curvatures, furthermore allowing us to estimate the liquid velocities as a function of rate of volume displaced. The liquid droplets at either end of the channel act as reservoirs, thus compensating for the liquid displaced by the vapor bubble. In addition, these droplets also ensure that the evaporation of the liquid to the ambient surrounding doesn't deplete the liquid within the channel. 

After attaining the capillary diameter ($d_\tu{h}$), the vapor bubble elongates along the axial (along $x$) direction with the cross-section as that of the channel. Thus, for an elongated bubble, when $X \ll l_\tu{L}$, the equation of motion of the liquid column within the channel is \cite{10.1017/S0022112009007381}
\begin{equation}
l_\tu{L} \rho_\tu{L} \dvtt{X} = p_\tu{V}(t) - p_{\infty} - \mathcal{R}\, \dvt{X},
\label{EOM}
\end{equation}
where $\rho_\tu{L}$ is the liquid density, $p_\tu{V}(t)$ is the pressure inside bubble, $p_{\infty}$ is the ambient pressure and $\mathcal{R}$ the hydraulic resistance of the channel. Due to strong confinement in two directions, the resulting flow is quasi-1D along the longitudinal axis $x$, which justifies the use of a one-dimensional model. The expressions of the steady state hydraulic resistance derived using laminar flow theory for different cross-sections are \cite{10.1039/9781849737067-00001}: $32\mu_\tu{L} l_\tu{L} / a^2$ (circle); $28.4\mu_\tu{L} l_\tu{L} / a^2$ (square) and $\approx 12\mu_\tu{L} l_\tu{L} / [(1-0.63b/a)b^2]$ (rectangle), where $\mu_\tu{L}$ is the dynamic viscosity of the liquid and $a,b$ are the cross-section's edge lengths with $b<a$. For circular and square cross-sections $d_\tu{h}=a$, and for a rectangular cross-section $d_\tu{h}=2ab/(a+b)$. By non-dimensionalizing Eq.\,\ref{EOM}, we obtain the characteristic velocity ($\upsilon$) and timescale ($\tau$) for the channel geometry as $\upsilon=(p_{\infty}-p_\tu{V}(t))/\mathcal{R}$ and $\tau=(l_\tu{L}\rho_\tu{L})/\mathcal{R}$, respectively. 

\begin{figure}[!t]
\centering
\includegraphics[width=0.8\linewidth]{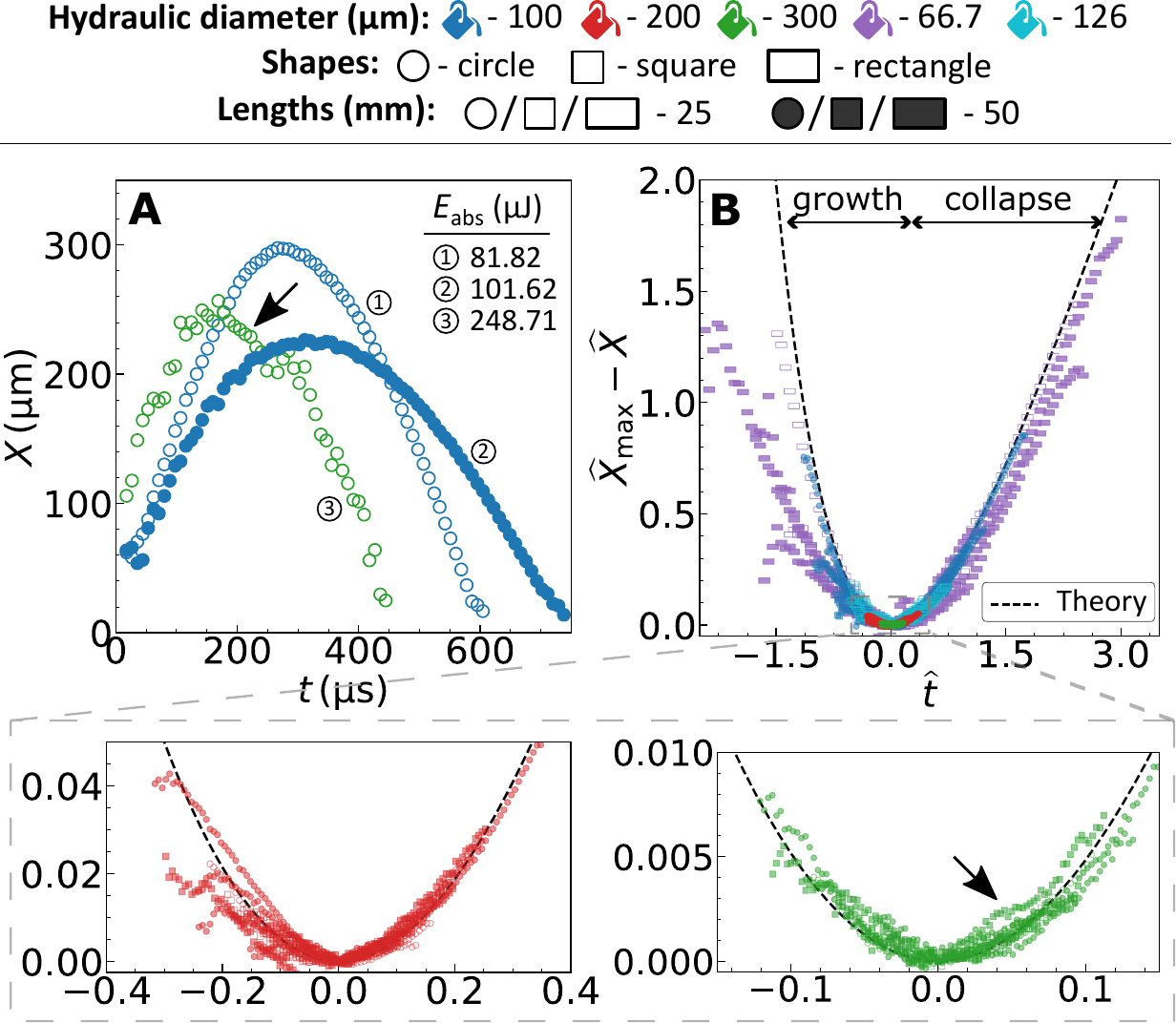}
\caption{A. Representative bubble dynamics for different channel geometries. B. Universal motion of bubbles within channels with different size, shape and length. The dashed line represents the developed theory, Eq.\,\ref{dimX}. The marker colors represent the hydraulic diameters ($d_\tu{h}$), the shapes represent the cross-section and the facecolor represent the lengths ($L$). The graphical marker symbols and colors established here are followed throughout this article. The black arrow represents the region of deviation(s) from the expected dynamics.}
\label{figure 2}
\end{figure}

Since $p_\tu{V}(t)$ is reported to be significant only during the first $\approx 10\%$ of the bubble's lifetime \cite{10.1016/S0017-9310(99)00027-7}, we simplify the problem by dropping this term ($p_\tu{V}(t)=0$). Thus, the liquid initially gains momentum from the impulse offered by the vapor pressure. Subsequently, the bubble continues to grow due to the liquid's inertia, even though the pressure inside the bubble is much less than the ambient pressure  \cite{10.1017/S0022112009007381}. At the end of the growth phase, the liquid momentum is fully dissipated by the resistance of the channel walls, following which the ambient pressure collapses the bubble. We describe the bubble growth and collapse by solving the ordinary differential equation (ODE) from Eq.\,\ref{EOM} and derive the following dimensionless equation for the dynamic bubble size:
\begin{equation}
\widehat{X}_\tu{max}-\widehat{X} = \widehat{t} + \exp(-\widehat{t}) - 1,
\label{dimX}
\end{equation}
where $\widehat{X}=X/(\upsilon\tau)$ and $\widehat{t}=(t-t_0)/\tau$ are the dimensionless bubble size and time, respectively. The ODE is solved using the boundary condition $X=X_\tu{max}$ at $t=t_0$, where $X_\tu{max}$ is the maximum bubble size. Fig.\,\ref{figure 2}A shows the representative bubble dynamics from experiments for different channel cross-sections and lengths. The bubble's maximum size, lifetime and the growth/collapse velocities vary significantly for different microchannel geometries. Fig.\,\ref{figure 2}B shows the experiments to follow the derived general expression for dynamic bubble size when non-dimensionalized using the characteristic scales (see \SI{1} \cite{SI} for unscaled results). 

In most cases, we see a deviation in the bubble size from theory in the first 10\% of the bubble's lifetime, as expected. This can be attributed to the significant vapor pressure inside the bubble due to instantaneous phase change during formation ($\sim$ 100 bar) which then decreases abruptly due to rapid change in the bubble volume \cite{10.1017/S0022112009007381}. Following the phase change, in the time range of $O(\upmu\tu{s})$ the vapor pressure is reported to be still very high $\sim 8\,\tu{bar}$ corresponding to a saturation temperature of $\sim 170\,^{\circ}$C \cite{10.1017/S0022112009007381}. After which the change in pressure and therefore the corresponding saturation temperature is rapid, 0.15\,bar/$\upmu$s or 2\,$^{\circ}$C/$\upmu$s approximately. Moreover, the force exerted by the channel wall during the radial ($r$, see Fig.\,\ref{figure 1}) growth of the bubble appearing during this first 10\% of the bubble's lifetime will also influence its dynamics \cite{10.1115/1.1824131}. Thus a more sophisticated theoretical model incorporating the rapid change in pressure/temperature, change in the bubble geometry from spherical to elongated, and wall forces in radial direction is necessary to accurately match the observed bubble dynamics in experiments. We therefore will overlook the accuracy within this time regime as its effect over the calculation of the characteristic parameters such as bubble lifetime, Womersley and Reynolds numbers are not large enough, discussed in detail in the following (sub)sections.

In Fig.\,\ref{figure 2}A, for $d_\tu{h}=300\,\upmu$m, we see a rapid acceleration during the start of the bubble collapse (pointed using a black arrow). Similarly, in Fig.\,\ref{figure 2}B, we also see a deviation from theory during the start of bubble collapse becoming more significant with increasing hydraulic diameter ($d_\tu{h}\geq200\,\upmu$m). We attribute these to instabilities, discussed in detail under the section ``Emergence of instabilities" later in this article.\\

\subsection*{Bubble lifetime} 
\indent While the dynamic size of the vapor bubble quantifies the liquid velocity or flow rate, the lifetime of the oscillating bubble governs the oscillatory time period of flow. In this work, we focus on the primary expansion and collapse of the bubble and ignore the bubble rebound caused by high pressures and temperatures \cite{10.1016/S0894-1777(02)00182-6}. Thus, the lifetime of the bubble is the time taken for one oscillation, $t_\tu{osc}$. In Eq.\,\ref{dimX}, by substituting $\widehat{X}=0$, we obtain the bubble's time of formation and collapse. The analytical prediction of the bubble's lifetime is
\begin{equation}
\widehat{t}_\tu{osc} = t_\tu{osc} / \tau = \tu{W}_0 (-e^{-\xi}) - \tu{W}_{-1} (-e^{-\xi}), 
\label{dimt}
\end{equation}
where W$_0$ is the principal branch of the Lambert $W$ function and W$_{-1}$ its only other real branch, and $\xi=1+\widehat{X}_\tu{max}$. The dimensionless times $\tu{W}_{-1} (-e^{-\xi}) + \xi$ and $\tu{W}_{0} (-e^{-\xi}) + \xi$ correspond to the time span of bubble expansion and collapse, respectively. We note that the dimensionless time in Figure\,\ref{figure 2}B is negative due to the choice of time zero at the maximum bubble size.  
\begin{figure}[!t]
\centering
\includegraphics[width=0.8\linewidth]{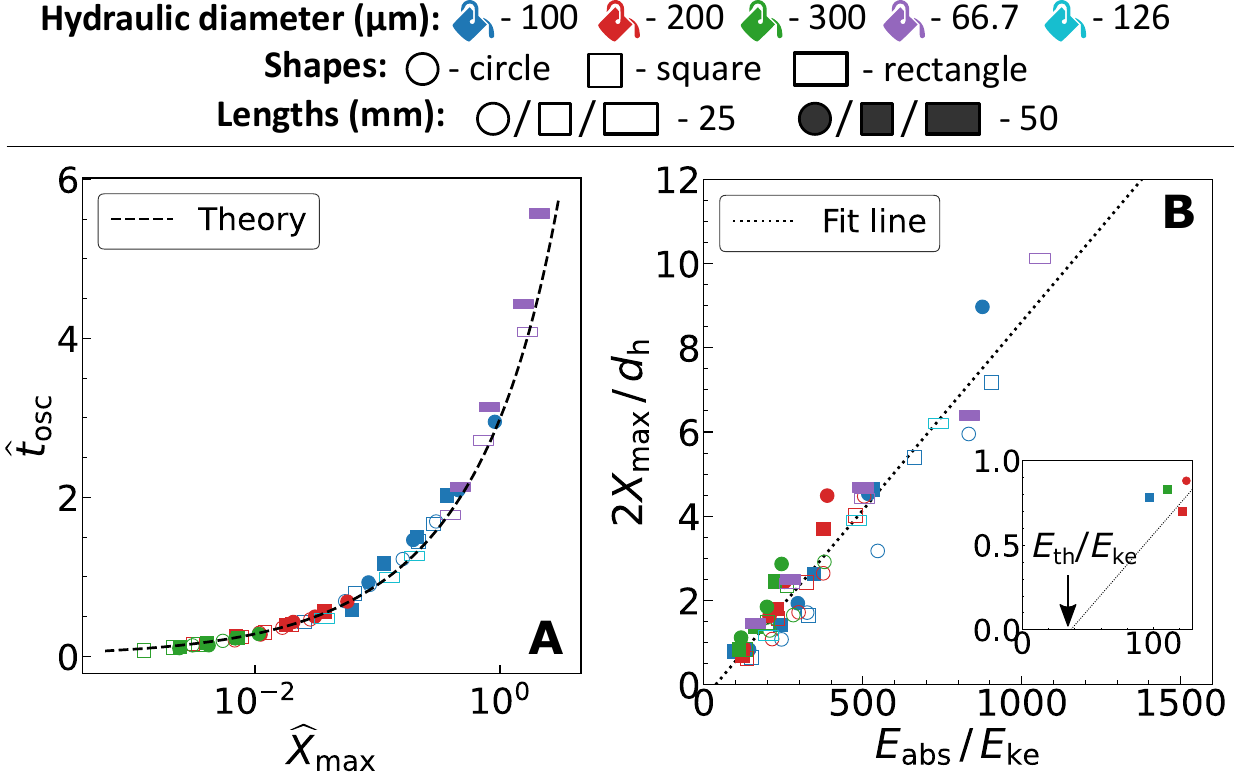}
\caption{A. Experimentally determined dimensionless lifetime ($\widehat{t}_\tu{osc}$) against maximum size ($\widehat{X}_\tu{max}$) of the bubble (markers) and comparison with the theoretical prediction from Eq.\,\ref{dimt} (dashed line). B. General linear relation between the non-dimensionalized experimentally determined maximum bubble size ($X_\tu{max}$) and laser energy absorbed by the liquid ($E_\tu{abs}$). The size and energy are non-dimensionalized using the channel hydraulic diameter ($d_\tu{h}$) and unitary kinetic energy ($E_\tu{ke}$) - calculated using the channel geometry and liquid properties. The inset illustrates the intercept in the plot which corresponds to the threshold laser energy for bubble formation ($E_\tu{th}$). The coefficient of determination for the fit line is $R^2=0.928$.}
\label{figure 3}
\end{figure}

Figure\,\ref{figure 3}A illustrates the relation between the dimensionless bubble lifetime and maximum bubble size. We report a good agreement between the experiments and the theoretical prediction from Eq.\,\ref{dimt}. See \SI{1} \cite{SI} for unscaled data presented in Fig.\,\ref{figure 3}A. There is a decrease in the bubble lifetime with an increase in the hydraulic diameter for a fixed channel length and $X_\tu{max}$. This is caused by the decrease in hydraulic resistance with increasing hydraulic diameter, resulting in faster bubble dynamics. While the exact solution to the Lambert $W$ function in Eq.\,\ref{dimt} accurately captures the experiments, an approximation of this equation is discussed in \SI{1} \cite{SI} to help solve the equation with more commonly used mathematical functions. The approximation however does decrease the accuracy of the results. 

The above sections discuss the theory for dynamic bubble size ($X(t)$, Eq.\,\ref{dimX}) and lifetime ($t_\tu{osc}$, Eq.\,\ref{dimt}). The proposed theory demands $X_\tu{max}$ as the only necessary parameter to compare with experiments. Thus it is necessary to have an estimate of $X_\tu{max}$ as a function of the laser energy supplied - a control parameter in experiments.\\ 

\subsection*{Energy balance}
\indent Figure\,\ref{figure 3}B represents a general scaling law using empirical correlation for the maximum bubble size ($X_\tu{max}$) against the absorbed laser energy, $E_\tu{abs}$. The parameters $X_\tu{max}$ and $E_\tu{abs}$ are non-dimensionalized with respect to the hydraulic diameter and kinetic energy of the liquid with unitary velocity, respectively. The liquid kinetic energy with unitary velocity ($E_\tu{ke}$) is calculated as
\begin{equation}
    E_\tu{ke} = A l_\tu{L} \rho_\tu{L},
\end{equation}
where $A$ is the cross-sectional area of the channel. This approach to determine $X_\tu{max}$ via empirical correlation avoids the need for a more sophisticated numerical model with energy balances and phase transition \cite{10.1016/S0894-1777(02)00182-6}. 

For a certain cross-sectional shape, the energy required to achieve a certain $X_\tu{max}$ increases with the hydraulic diameter and length of the channel (see \SI{1} \cite{SI} for unscaled results of Fig.\,\ref{figure 3}B). In Fig.\,\ref{figure 3}B we see the $E_\tu{abs}$ to be at least two orders of magnitude higher than the $E_\tu{ke}$. This observation is in agreement with the numerical simulations by Sun \textit{et al.} \cite{10.1017/S0022112009007381}, where most of the absorbed laser energy heats up the liquid directly surrounding the bubble, rather than transforming into the kinetic energy of the liquid.

In the experiments, the amount of laser energy absorbed by the liquid depends on the liquid properties and channel geometry. Based on the Beer-Lambert law \cite{beerlambertlaw}, the absorption coefficient of the liquid and the distance the light travels through the liquid is used to estimate the amount of light absorbed, $E_\tu{abs}$. Furthermore, the channel's wall thickness, material, cross-section shape and dimension can influence the laser energy available for the liquid to absorb. Thus the energy absorbed by the liquid in experiments was calibrated by measuring the difference of energy transmitted with the microchannel containing water and water-dye mixture (working fluid). Since water has very low absorption coefficient ($<0.001\,\tu{cm}^{-1}$) for the laser wavelength used (532\,nm) \cite{10.1364/AO.36.008699}, we attribute the measured energy difference to the laser energy absorbed by the dye.

For rectangular channels, the longer edge length was used for supporting the channel over the stage (see Fig.\,\ref{figure 1}), while the laser light was passed along the shorter edge. The details on the energy absorption measurement technique can be found in \SI{2} \cite{SI}. The estimated absorption coefficient of liquid, $\alpha$, is $86.54\,\tu{cm}^{-1}$. This estimated value is different from measurements from the spectrometer (DR6000, Hach), 173\,cm$^{-1}$, which employs an unfocused and continuous light source. We attribute this difference in absorption coefficient to the non-linear absorbance (saturable absorption) of the liquid due to high energy intensities ($\sim \tu{GW} \tu{cm}^{-2}$) as the laser is focused and has short laser pulse duration (4\,ns).

The fit in Fig.\,\ref{figure 3}B has an intercept for $E_\tu{abs}$, implying there exists a threshold absorbed energy ($E_\tu{th}$) only above which a vapor bubble forms. For a bubble to form, theoretically we use the spinodal temperature of water as the necessary condition \cite{10.1007/s10953-009-9417-0} - the temperature at which water explosively turns into vapor. Thus, a sensible heat corresponding to $\Delta T = 279\,\tu{K}$ (at 1\,atm and with respect to room temperature of 298\,K) and a latent heat proportional to enthalpy of vaporization ($H_\tu{L}$) is minimum required at the laser spot for bubble formation. \SI{2} \cite{SI} provides the derivation for analytical expression of $E_\tu{th}$ based on the energy balance for liquid. The threshold absorbed laser energy is, 
\begin{equation}
    E_\tu{th} = (1-10^{-\alpha d_\tu{L}})\frac{\pi w_0^2 \rho_\tu{L} (c_p\Delta T + H_\tu{L})} {\alpha\log(10) 10^{-\alpha (d_\tu{L}/2)}}.
    \label{Eabs}
\end{equation}
Where $w_0$ is the laser spot radius, $c_p$ is the specific heat and $d_\tu{L}$ is the distance the light will travel through the liquid. $d_\tu{L}$ is equal to the hydraulic diameter ($d_\tu{h}$) for circular and square channels. For rectangles it is the shorter edge length of the cross-section. The parameter values for the calculation of Eq.\,\ref{Eabs} are presented in Table\,\ref{parameter values}.
\begin{table}[h]
\centering
\caption{Parameter values for theoretical calculation of the threshold energy for bubble formation using Eq.\,\ref{Eabs}.}
\begin{tabular}{>{\centering\arraybackslash}p{7em}>{\centering\arraybackslash}p{16em} >{\centering\arraybackslash}p{20em}} \hline
Parameter & Description & Value  \\
\midrule
$\Delta T$ & temperature rise & 279\,K (with respect to room temperature of 298\,K) \\
$c_p$ & specific heat & $4200\, \tu{J/(kg\,K)}$ \\
$H_\tu{L}$ & latent heat of vaporization & $1377.6\,\tu{kJ/kg}$ (at 577\,K)\\
$\rho_\tu{L}$ & liquid density & $1000\,\tu{kg/m}^3$ \\
$\alpha$ & liquid absorption coefficient & $86.54 \,\tu{cm}^{-1}$ (measured experimentally, see \SI{2} \cite{SI}) \\
\bottomrule
\end{tabular}
\label{parameter values}
\end{table}

\begin{figure}[h]
\centering
\includegraphics[width=0.7\linewidth]{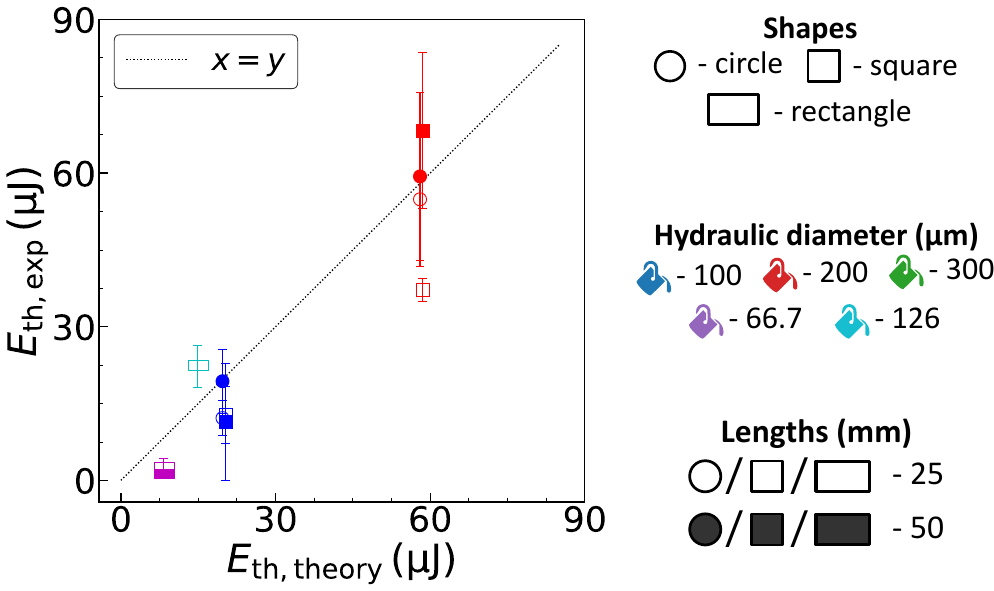}
\caption{The threshold laser energy absorbed for bubble formation estimated from experiments ($E_\tu{th,exp}$) against theory ($E_\tu{th,theory}$) presented in Eq.\,\ref{Eabs}.}
\label{figure threshold}
\end{figure}

In Eq.\,\ref{Eabs}, the other necessary parameter for calculation is the laser spot diameter, $2w_0$. Theoretically, the laser spot diameter is calculated using the expression $2w_0 = 4M^2\lambda f / (\pi D_\tu{L}) = 1.69\,\upmu$m. Where $\lambda=532\,\tu{nm}$ is the laser wavelength, $f=10\,\tu{mm}$ is the objective lens focal length, $D_\tu{L}=4\,\tu{mm}$ is the laser beam diameter and the beam quality parameter $M^2=1$ (assuming a perfect Gaussian profile). The depth of field (DOF) of the focused laser beam is $2\pi w_0^2 / (M^2 \lambda) = 8.47\,\upmu$m. However, these theoretical calculation for spot size and DOF can be different in experiments due to the optical aberrations caused by the channel walls \cite{10.1364/AO.38.003636}. From Fig.\,\ref{figure threshold}, we see a good agreement between experiments ($E_\tu{th,exp}$) and theory ($E_\tu{th,theory}$) presented in Eq.\,\ref{Eabs} for the threshold laser energy absorbed by the liquid for bubble formation. The laser spot diameter used in the analytical expression is $9\,\upmu$m, which is estimated based on the best fit of the theory to the experiments. This value of the laser spot diameter is same order of magnitude as the theoretically calculated value, thus emphasizing its validity. Furthermore, in Fig.\,\ref{figure threshold} we see a deviation in the threshold laser energy absorbed. We attribute the deviation to the laser aberrations due to channel wall curvature and thickness that can influence the laser spot diameter and therefore the threshold laser energy. In addition, the $r$ position of the laser spot can also be affected by the confinement wall curvature and thickness, resulting in a larger $E_\tu{th}$ as the laser spot moves towards $r=d_\tu{L}/2$. The channel specifications used in this work are summarized in Table\,\ref{channel geometry}.

The above analysis thus provides us with an expression for bubble formation as a function of the energy absorbed by the liquid - a parameter that can be measured in experiments. The estimation of the laser energy threshold is one of the key design parameters necessary to choose the range of the laser energy required based on the channel geometry, liquid properties and objective lens specification.\\

\subsection*{Emergence of instabilities} 
\begin{figure}[!hp]
\centering
\includegraphics[width=0.9\linewidth]{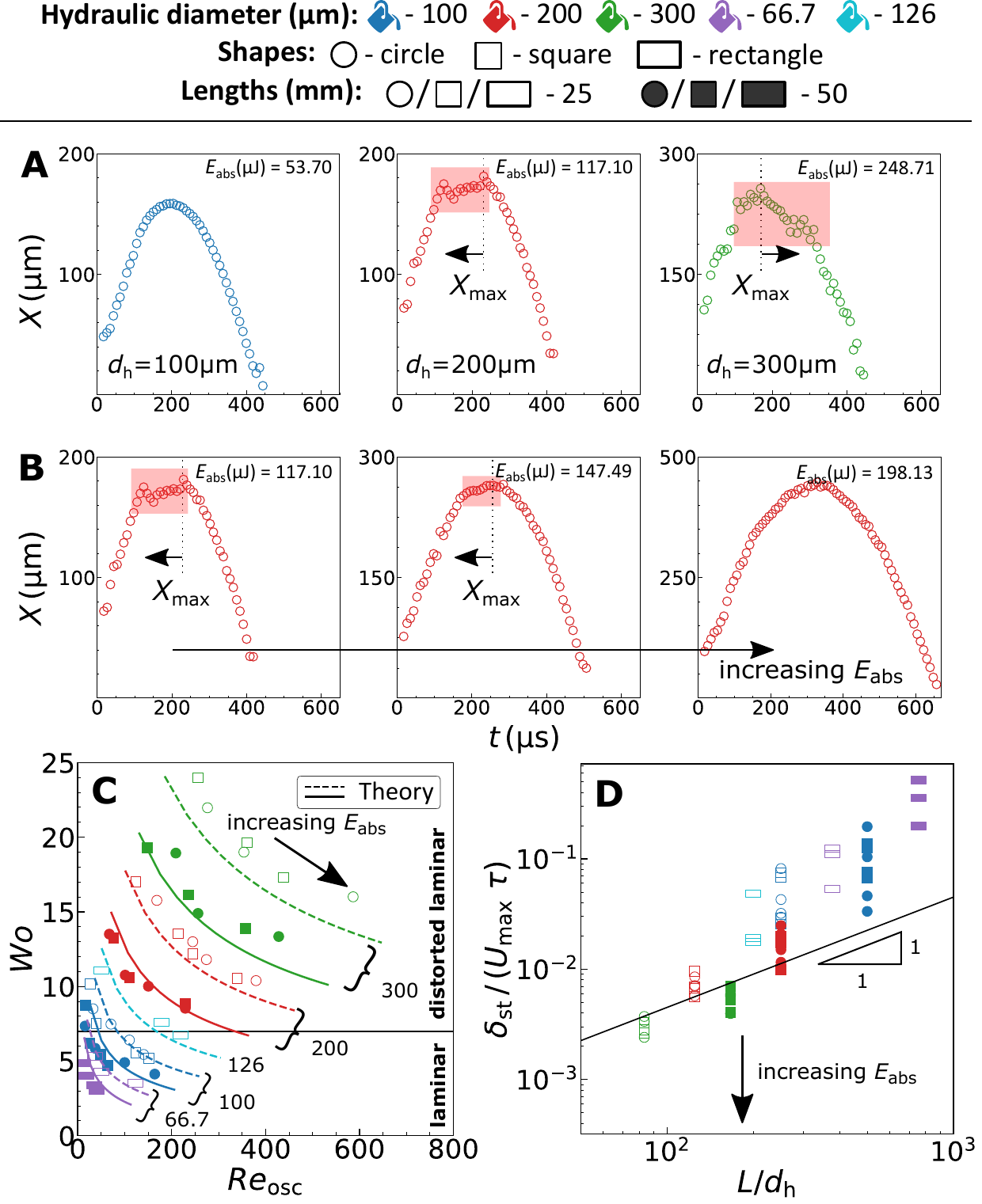}
\caption{A,B. Representative dynamic bubble size curves illustrating the emergence of instabilities. The zones of the instabilities are highlighted using a shaded rectangular area. The arrows represent if the instabilities occur before or after $X_\tu{max}$. A. Illustrates the experimental data for different $d_\tu{h}$ with similar oscillation time. The instabilities emerge with increasing $d_\tu{h}$. B. Illustrates the data for $d_\tu{h}=200\,\upmu$m with increasing laser energies. The instabilities disappear with increasing $E_\tu{abs}$. C. Flow stability diagram with the transition line at $\textit{Wo}=7$ \cite{10.1299/jsme1958.25.365}. The markers represent the experiments and the lines represent the analytical estimate. The numbers correspond to the channel hydraulic diameters (in $\upmu$m) with the dashed and solid lines representing the channel lengths L = 25 and 50\,mm, respectively. D. The dimensionless convective timescale against the $L/d_\tu{h}$ aspect ratio. The partition line is a linear relation between the $x$ and $y$ axes with $45\times10^{-6}$ as the slope and the origin as the intercept.}
\label{figure 4}
\end{figure}

\indent In Fig.\,\ref{figure 4}A, for $d_\tu{h}\geq200\,\upmu\tu{m}$ we see the bubble dynamics deviating from an expected bell curve like profile. Interestingly, the deviation occurs before $X_\tu{max}$ for $d_\tu{h}=200\,\upmu\tu{m}$ and after $X_\tu{max}$ for $d_\tu{h}=300\,\upmu\tu{m}$. Furthermore, Fig.\,\ref{figure 4}B shows that for $d_\tu{h}=200\,\upmu\tu{m}$ the deviation disappears for larger $X_\tu{max}$. We hypothesise the observed deviations to the boundary layer instabilities near channel walls. These instabilities emerge due to the oscillating nature of the flow and unlike unidirectional flows it cannot be characterized only using the Reynolds number. Thus, the flow instability in oscillating channel flow is characterized using a peak oscillatory Reynolds number $Re_\tu{osc} = \rho_\tu{L}U_\tu{max}d_\tu{h}/\mu_\tu{L}$ and a Womersley number $\textit{Wo} = d_\tu{h}/2\sqrt{\omega\rho_\tu{L}/ \mu_\tu{L}}$ \cite{10.1017/S0022112076000177,10.3390/en14051410}. $U_\tu{max}$ is the maximum mean flow velocity during the bubble growth, which on average occurs in the middle of the duration of growth, and $\omega = 2\pi/t_\tu{osc}$ is the angular frequency of oscillation. Fig.\,\ref{figure 4}C shows these dimensionless numbers for the experiments performed in this work.

A simple approach to estimate the flow velocity, $U$, using theory can be formulated by using the analytical expression for $t_\tu{osc}$ (Eq.\,\ref{dimt}) and differentiating Eq.\,\ref{dimX} with respect to time. The resulting dimensionless flow velocity is
\begin{equation}
U/v = -1 + \exp(-\widehat{t}).
\label{dimU}
\end{equation}
By substituting half the bubble growth time $[\tu{W}_{-1} (-e^{-\xi}) + \xi]/2$ for $\widehat{t}$, we calculate a corresponding maximum mean flow velocity $U_\tu{max}$ and $Re_\tu{osc}$.

The lines in Fig.\,\ref{figure 4}C represent the analytical prediction of the dimensionless flow parameters for varying $X_\tu{max}$ in all channel geometries. We report a good agreement between the experiments and the analytical estimate. The horizontal line in the figure, $\textit{Wo}=7$, represents the empirically observed laminar to distorted laminar transition value from literature for a perfectly oscillating flow with sinusoidal pressure gradient \cite{10.1299/jsme1958.25.365}. To predict the transition, the onset and growth of instability is determined using the parameters: (i) the position of the point of inflection, and (ii) the convective timescale, respectively. Das \textit{et al.} \cite{10.1017/S0022112098002572}, using pulsatile flow theory with a sinusoidal mean velocity profile \cite{10.1113/jphysiol.1955.sp005276}, showed the dimensionless position of the point of inflection to be independent of the \textit{Wo} above a critical Stokes parameter, $d_\tu{h}/(2\delta_\tu{st}) = \textit{Wo}/\sqrt{2} \approx 3.6$, where $\delta_\tu{st} = \sqrt{2\mu_\tu{L}/(\rho_\tu{L}\omega)}$ is the Stokes layer thickness. This theoretical critical value of $\textit{Wo}\approx5$ is below the empirically observed critical transition value of $\textit{Wo}=7$. Thus, the aforementioned analysis of instability using $\delta_\tu{st}$ demonstrates the critical $\textit{Wo}$ in flow transition to be independent of the position of the point of inflection. However, the source of instability can still be attributed to the inflection of the velocity profiles near channel walls occurring due to the flow reversal (boundary layer separation) during the deceleration phase of the liquid. To further emphasize the observed deviation in Fig.\,\ref{figure 4}A as a cause of instability and not due to assumptions underlying the theory under consideration, we compare it to laminar flow due to an oscillating pressure gradient. Just before the bubble attains its maximum size, detachment of the boundary layer at the confining walls has been reported for flows between parallel plates based on numerical analysis \cite{10.1103/PhysRevFluids.2.014301}. The flow detachment exists due to the mismatch between the direction of the pressure gradient ($\partial p/\partial x > 0$) and fluid flow ($U>0$). In accordance, our estimated flow profiles using theory with oscillating pressure gradient for microchannels \cite{10.1007/s11242-016-0652-8} also predict a flow reversal occurring close to the channel walls (see \SI{3} \cite{SI}), similar to others \cite{10.1007/s00348-009-0781-8}. However, the flow reversal near walls should decelerate the flow as the bubble begins to collapse - opposite to what is observed in experiments (in Fig.\,\ref{figure 4}A, $d_\tu{h}=300\,\upmu$m). This is a consequence of the instabilities causing flow distortion.

While the inflection in velocity profile near channel walls can lead to the onset of instabilities, the growth and therefore the time of emergence of instabilities is governed by the convective timescale. Fig\,\ref{figure 4}D shows the dimensionless convective timescale from experiments, $\delta_\tu{st}/(U_\tu{max}\tau)$, for different $L/d_\tu{h}$ ratios. Small characteristic timescales of the channels ($\tau$) compared to convective timescale ($\delta_\tu{st}/U_\tu{max}$) would correspond to large resistance to flow offered by the channel walls as $\tau \propto 1/\mathcal{R}$. Large resistance ($\mathcal{R}$) in other words would mean a viscosity-dominated flow that effectively would dampen any perturbations/instabilities. However, as $\tau$ increases the instability would develop during the deceleration phase and prevail in the acceleration phase \cite{10.1017/S0022112098002572}. This argument on instability evolution explains the observed orientation of flow distortion relative to $X_\tu{max}$ in Fig\,\ref{figure 4}A, when analyzed using the partition line depicted in Fig\,\ref{figure 4}D. The partition line is adapted based on the experiments from this work which emphasizes the time of emergence of instabilities. As a data point approaches the partition line, the instabilities die out due to the motion-less state (zero velocities) at $X_\tu{max}$, while below the partition line the instabilities sustain and prevail due to larger kinetic energies ($\propto U_\tu{max}$).

In summary, the observed deviations in the dynamic bubble size from a bell curve like profile can be attributed to the disruption of the momentum boundary layer near channel walls \cite{10.1115/1.1490375} that would alter the channel hydraulic resistance and hence the mean velocities. However, these distorted laminar states are transient and therefore revert (decay) to laminar flow over time \cite{10.1115/1.1490375,10.1017/S0022112076000177}. Consequently, there exists a momentary deviation in dynamic bubble size as seen from the representative images in Fig.\,\ref{figure 4}A.\\

\noindent \textbf{\large Conclusions}\\
\indent By combining experiments and theory, we demonstrated universal behavior of fluid flows and transitions caused by transient laser-induced bubbles within microchannels of different geometries. This generalized approach to flow characteristics will aid the optimization of channel design and laser energy based on the application specific functionality. Contrary to the general assumption of undistorted laminar flow due to low Reynolds numbers ($Re<1000$) in bubble-powered micro-systems, we report flow instability. The instability originates due to the oscillating nature of the flow when the boundary layer separation occurs near channel walls during flow deceleration. During deceleration, the inflection point in the velocity profile near channel walls becomes unstable for larger flow oscillation frequencies ($\textit{Wo}>7$). Following the onset of instability, the time of emergence as distorted laminar states is discussed using the convective timescale associated with the flow. However, the distorted laminar states due to the growth of instability are transient and therefore decay rapidly compared to the overall flow oscillation time.

While this work is first to argue the existence of flow distortion in cavitation actuated flows in microfluidic channels, the measurement of such distortions in piston actuated flows in channels with cross-sectional edge length of $O(10\,\tu{mm})$ have been extensively documented in the literature. The adapted measurement techniques in literature involve hot-wire anemometry \cite{10.1017/S0022112075001826,10.1299/jsme1958.25.365,10.1017/S0022112076000177,10.1017/S002211209100112X}, laser doppler anemometry \cite{10.1017/S002211209100112X,10.1017/S0022112080001267,10.1017/S0022112091002100,10.1017/S0022112098002559}, particle image velocimetry \cite{inproceedings} and flow visualization using dyes \cite{10.1017/S0022112098002572} for liquids, while for gases the techniques involve the use of transpiring walls \cite{10.1115/1.1490375} and smoke \cite{10.1017/S0022112075001826} for flow visualization. This work therefore does not employ any dedicated measurement technique to study flow distortion. However the argument on flow distortion is supported by characterizing the flow using the dimensionless numbers: Reynolds ($Re$) and Womersley ($W\!o$) numbers, and order of magnitude analysis using convective time of the flow $\delta_\tu{st}/(U_\tu{max} \tau)$. In future, for studies that might examine the flow distortion in detail in our proposed system, we recommend the use of laser doppler anemometry, micro-particle image velocimetry and dyes since the presence of the hot-wire probe can promote flow distortion \cite{10.1017/S002211209100112X}.

We note that this study may also be relevant to explain a new phenomenon in the field of vesicle/cell deformation due to flow instability, challenging prior understanding based on resonance \cite{10.1039/C3SM51399H} and undistorted laminar shear \cite{10.1103/PhysRevFluids.2.014301,10.1038/nature01613}. Thus our findings inform rational design of oscillatory pulsatile flows in engineering systems with potential applications to cavitation powered actuators \cite{10.1039/D2LC00169A,10.1007/978-981-287-470-2_6-1,10.1038/srep09942,10.1038/srep04787}, confinement-induced acoustic cavitation \cite{10.1103/PhysRevApplied.14.024041} and biomimetic micromachines \cite{10.1002/adfm.202214130}.


Our work employs a simplified Darcy-Brinkman equation \cite{10.1007/s11242-016-0652-8} for pure fluids with transient flow equations assuming steady-state hydraulic resistance. Therefore, by additionally incorporating the Darcy number, effective viscosity and porosity \cite{10.1063/1.2792323}, this work can conveniently be extended to study flows within porous ducts with application to heat transfer in refrigeration \cite{10.1016/j.applthermaleng.2013.01.020} and biology \cite{10.1016/S0017-9310(03)00301-6}.\\

\section*{Data Availability}
\indent All the post-processed data that supports this study are provided in the supplementary information. The raw data that supports this study are available with the corresponding author. The data will be shared upon reasonable request.\\

\bibliography{references}

\section*{Acknowledgements}
\indent We thank Dr. Carlas S. Smith,  Ir. Aswin Raghunathan and Shih-Te Hung for their support with the design of the experimental setup. Thanks to Ir. Suriya Prakash Senthil Kumar for support with the image processing. Thanks to Dr. Mathieu Pourquie for the productive discussions. This work was funded by LightX project under NWO Open Technology Programme (project number 16714).\\

\section*{Author contributions statement}
\indent N.N. and H.B.E. designed and performed the experiments;
N.N., J.T.P. and R.H. analyzed the theoretical model;
N.N., J.T.P., R.H. and H.B.E. wrote the manuscript;
N.N., V.K. and D.I. developed the experimental setup;
N.N. and V.K. formulated the experimental methodology;
J.W. provided theoretical expertise.\\

\section*{Additional information}
The authors declare no conflict of interest.

\end{document}


\title{Unified Framework for Laser-induced Transient Bubble Dynamics within Microchannels} 

\author{Nagaraj Nagalingam}
\affiliation{Process \& Energy Department, Delft University of Technology, Leeghwaterstraat 39, 2628 CB Delft, Netherlands}
\author{Vikram Korede}
\affiliation{Process \& Energy Department, Delft University of Technology, Leeghwaterstraat 39, 2628 CB Delft, Netherlands}
\author{Daniel Irimia}
\affiliation{Process \& Energy Department, Delft University of Technology, Leeghwaterstraat 39, 2628 CB Delft, Netherlands}
\author{Jerry Westerweel}
\affiliation{Process \& Energy Department, Delft University of Technology, Leeghwaterstraat 39, 2628 CB Delft, Netherlands}
\author{Johan T. Padding}
\affiliation{Process \& Energy Department, Delft University of Technology, Leeghwaterstraat 39, 2628 CB Delft, Netherlands}
\author{Remco Hartkamp}
\affiliation{Process \& Energy Department, Delft University of Technology, Leeghwaterstraat 39, 2628 CB Delft, Netherlands}
\author{Hüseyin Burak Eral}
\email{h.b.eral@tudelft.nl}
\affiliation{Process \& Energy Department, Delft University of Technology, Leeghwaterstraat 39, 2628 CB Delft, Netherlands}

             

\maketitle

\tableofcontents

\section{Dynamic bubble size, bubble lifetime and bubble energy}

\begin{figure}[h]
\centering
\includegraphics[width=0.95\textwidth]{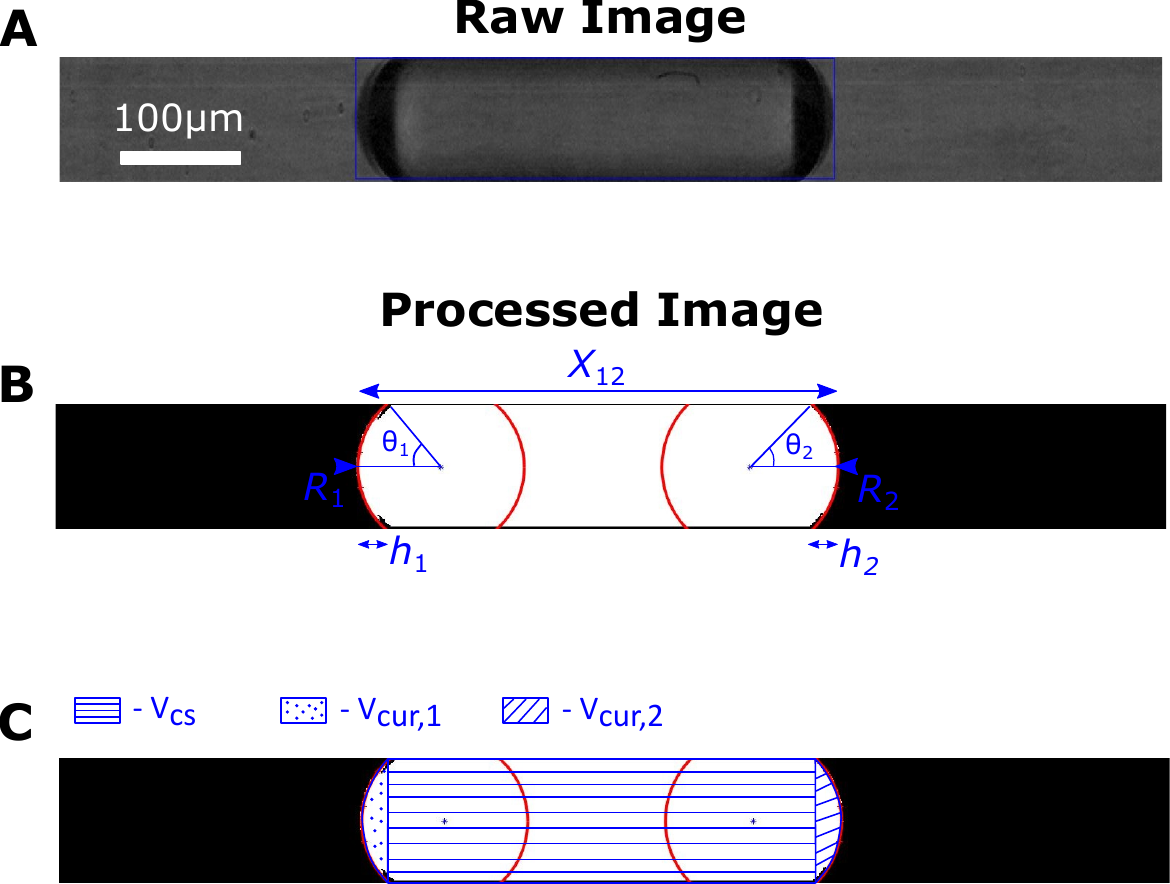}
\caption{(A) Raw image from experiments. (B) Processed image. $X_{12}$ is the distance between the extreme bubble ends. $R_1$ and $R_2$ are the radius of curvature for the left and right ends respectively. $h_1$ and $h_2$ are the width of the curvature for the left and right ends respectively. (C) $V_\tu{cs}$ is the cross-section volume and $V_\tu{cur}$ is the curvature volume. The total volume of the vapor bubble, $V_\tu{b} = V_\tu{cs} + 2 V_\tu{cur}$.}
\label{Image processing}
\end{figure}

\textbf{Determining $\bm{X}$.} Fig.\ref{Image processing} illustrates the image processing technique adapted to determine the bubble size. To eliminate the effects of curvatures on the estimation of $X$, we first calculate the bubble volume. The total volume of a vapor bubble, $V_\tu{b} = V_\tu{cs} + V_\tu{cur,1} + V_\tu{cur,2}$, can be expressed using the summation of the portions as expressed in Fig.\,\ref{Image processing}. Where $V_\tu{cs}$ is the cross-section volume and $V_\tu{cur}$ is the curvature volume. The effective bubble length incorporating the curvature effects over the bubble ends is $X=V_\tu{b}/A$.\\\\ 
The volume $V_\tu{cs} = A (X_{12} - h_1 - h_2)$, where A is the cross-section area of the channel. $A = \pi d_\tu{h}^2/4$ for circular cross-section, and $A = a \, b$ for square/rectangular cross-section. $d_\tu{h}$ is the channel's hydraulic diameter and, $a$ and $b$ are the cross-section's edge lengths with $b\leq a$. The calculation of volume $V_\tu{cur}$ requires the consideration of the curvature.\\

\noindent \textit{Circular cross-section}\\
\indent The bubble volume is 
\begin{equation}
    V_\tu{b} = A\, (X_{12}-h_1-h_2) + \sum_{i=1}^2 \, \pi h_i\,(3 d_\tu{h}^2/4 + h_i^2) / 6,
    \label{Vb}
\end{equation}
where,
\begin{equation}
h_i = R_i\,(1-\cos(\theta_i)) \hspace{1em} \textup{and} \hspace{1em} \theta_i = \sin^{-1}(d_\tu{h}/(2R_i)).
\end{equation}
Refer to Fig.\ref{Image processing} for all the other notations discussed in the section.\\

\noindent \textit{Square/Rectangle}\\
\indent The volumetric flow rate due to a curvature can be expressed as $\dot{V}_\tu{cur} = A U_\tu{mean,cur}$, where $U_\tu{mean,cur}$ is the average velocity.  The average velocity for a rectangular channel is calculated as \cite{10.1007/s11242-016-0652-8},
\begin{equation}
    U_\tu{mean,cur} = \frac{a^2 \Delta p}{4l_\tu{L}\mu_\tu{L}}\frac{\sin(\alpha_n)\beta_n}{A\!R \,\alpha_n^3} \left[A\!R - \frac{\tanh(\alpha_n A\!R)}{\alpha_n}\right].
    \label{Umean}
\end{equation}
Where $\alpha_n = \left(n-\frac{1}{2}\right) \pi$, $\beta_n = \frac{2(-1)^{n+1}}{\alpha_n}$ and $A\!R=b/a$. Furthermore, the velocity at the channel axis ($r=0$) is \cite{10.1007/s11242-016-0652-8},
\begin{equation}
    U_{r=0} = \frac{a^2 \Delta p}{4l_\tu{L}\mu_\tu{L}} \sum_{n=1}^\infty \frac{\beta_n}{\alpha_n^2} \left[1 - \frac{1}{\cosh(\alpha_n A\!R)}\right].
    \label{Ur=0}
\end{equation}

For square channels, $A\!R=1$, since $a=b$. Furthermore, Eqs.\,\ref{Umean}\,and\,\ref{Ur=0} can be solved to obtain the velocity ratio, $U_\tu{ratio} = U_\tu{mean,cur} / U_{r=0} = 0.477$, which is independent of the channel's hydraulic diameter or the cross-sectional edge lengths. Thus $\dot{V}_\tu{cur}$ can be calculated as, $\dot{V}_\tu{cur} = 0.477 A\, U_{r=0}$. Since $a=b=d_\tu{h}$, the cross-section area is $A=d_\tu{h}^2$. Analogous to $\dot{V}_\tu{cur}$, the $V_\tu{cur}$ is calculated by substituting $U_\tu{r=0}$ with $h_i$. We get, $V_\tu{cur,i} = 0.477\, h_i d_\tu{h}^2$. Thus for square channels,
\begin{equation}
    V_\tu{b} = A\, (X_{12}-h_1-h_2) + \sum_{i=1}^2 \, 0.477\, h_i d_\tu{h}^2.
\end{equation}

In this work, we employ two rectangular channels with $A\!R=0.5\,\,\tu{and}\,\,0.2667$ for $50\times 100$ (in $\upmu$m) and $80\times 300$ (in $\upmu$m) channels, respectively. Below we present the results for rectangular channels similar to square:
\begin{gather}
    V_\tu{b} = A\, (X_{12}-h_1-h_2) + \sum_{i=1}^2 \, 0.502\, h_i a\, b \hspace{1em} \tu{for $d_\tu{h}=66.7\,\upmu$m, and}\\
    V_\tu{b} = A\, (X_{12}-h_1-h_2) + \sum_{i=1}^2 \, 0.558\, h_i a\, b \hspace{1em} \tu{for $d_\tu{h}=126\,\upmu$m}.
\end{gather}

\begin{figure}[h]
\centering
\includegraphics[width=0.85\textwidth]{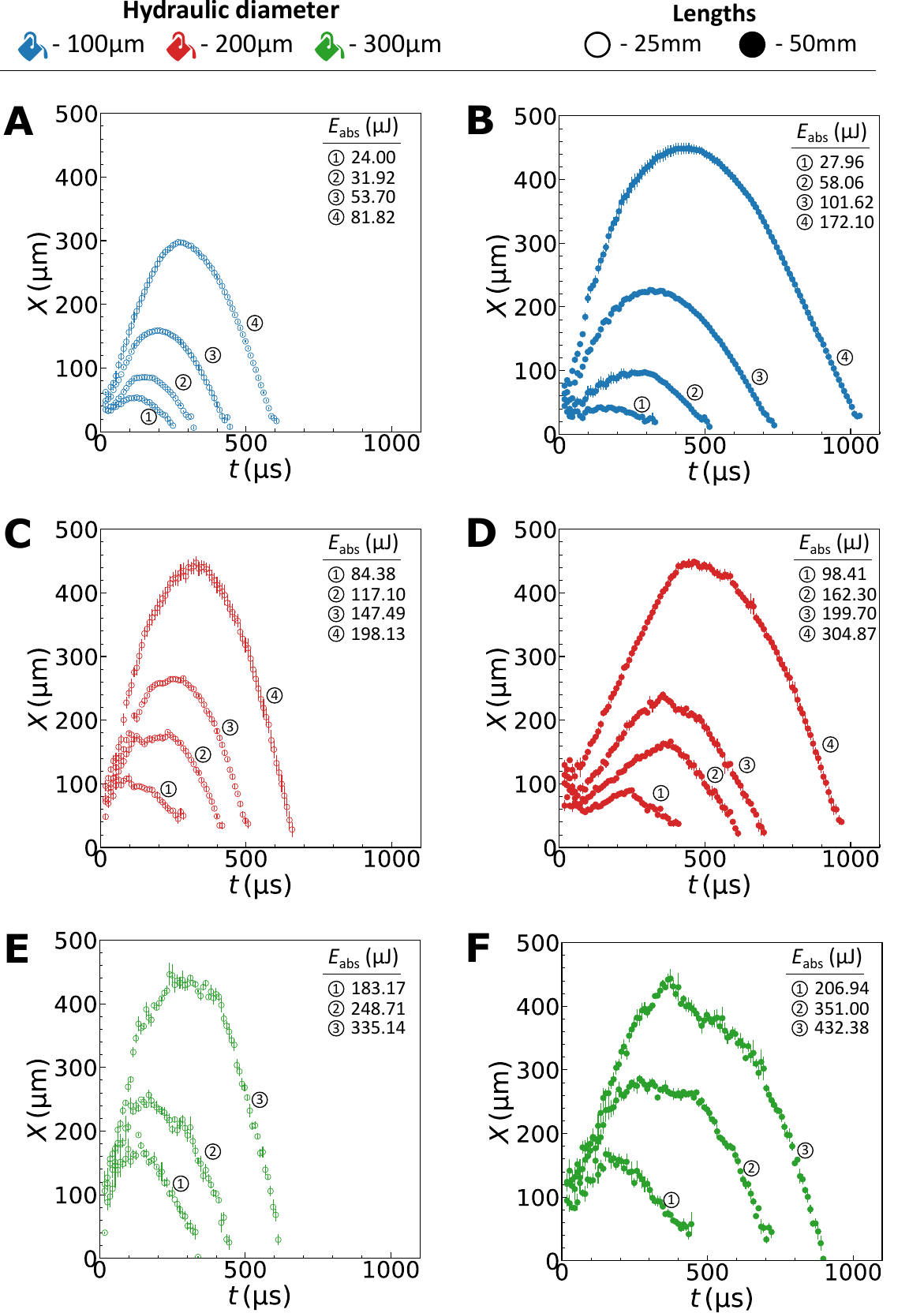}
\caption{(A-F) Bubble size ($X$) against time ($t$) for a channel with circular cross-section. The colors represent the dimension; the markerfacecolors represent the channel length. The error bars represent the standard error over 5 trials.}
\label{Xcircle}
\end{figure}

\begin{figure}[h]
\centering
\includegraphics[width=0.90\textwidth]{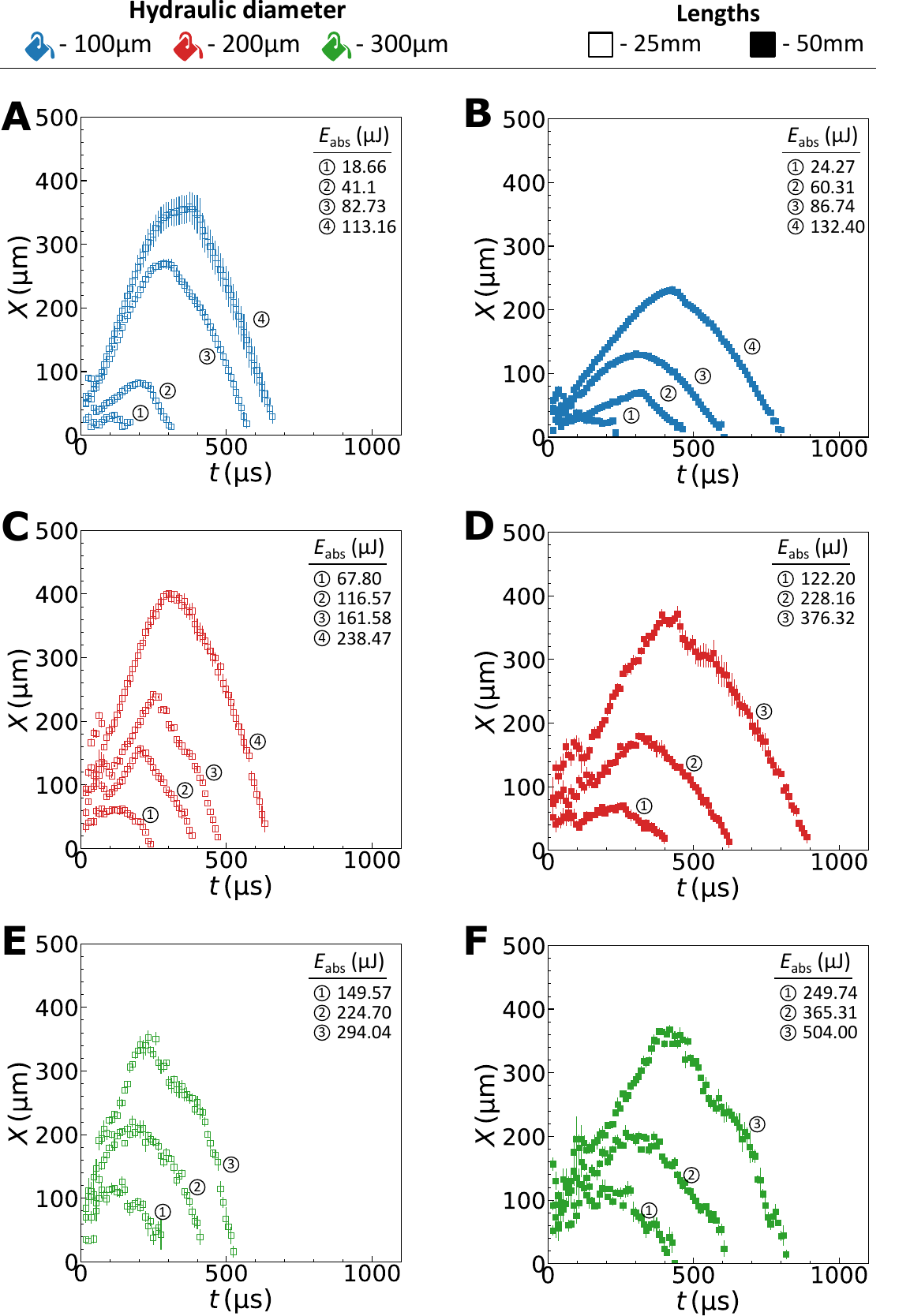}
\caption{(A-F) Bubble size ($X$) against time ($t$) for a channel with square cross-section. The colors represent the dimension; the markerfacecolors represent the channel length. The error bars represent the standard error over 5 trials.}
\label{Xsquare}
\end{figure}

\begin{figure}[h]
\centering
\includegraphics[width=0.90\textwidth]{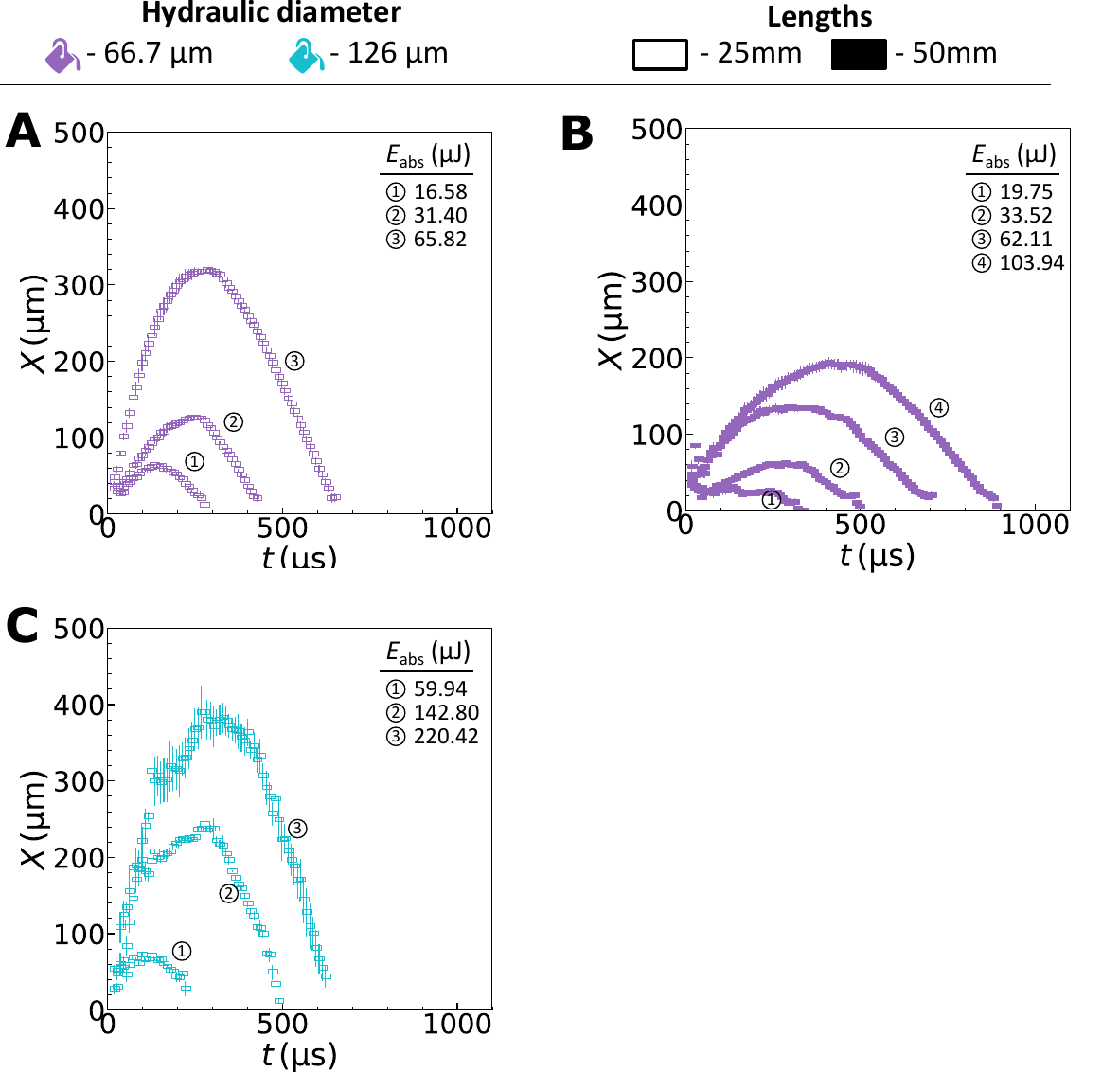}
\caption{(A-C) Bubble size ($X$) against time ($t$) for a channel with rectangular cross-section. The colors represent the dimension; the markerfacecolors represent the channel length. The error bars represent the standard error over 5 trials.}
\label{Xrectangle}
\end{figure}

\begin{figure}[h]
\centering
\includegraphics[width=0.95\textwidth]{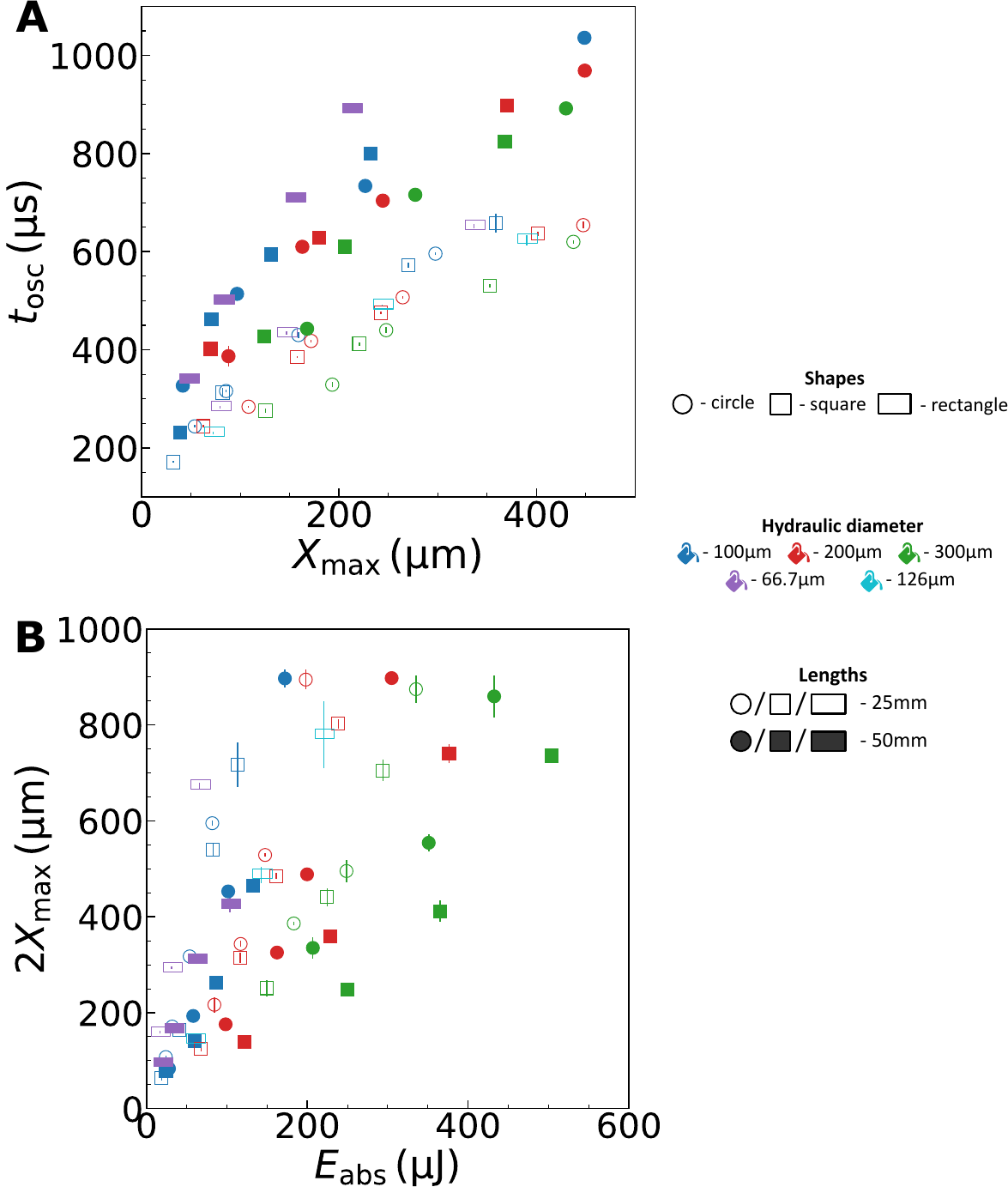}
\caption{(A) The lifetime ($t_\tu{osc}$) against maximum size ($X_\tu{max}$) of the bubble. (B) The maximum bubble size ($X_\tu{max}$) against the laser energy absorbed by the liquid ($E_\tu{abs}$). The marker shapes represent the shape of the cross-section; the colors represent the dimension; the markerfacecolors represent the channel length. The error bars represent the standard error over 5 trials.}
\label{time and energy}
\end{figure}

\textbf{Bubble lifetime and energy.} The non-dimensionalized plots (unscaled) for dynamic bubble size are represented in Figs.\,\ref{Xcircle}-\ref{Xrectangle}. Similarly, the non-dimensionalized plots (unscaled) for bubble lifetime and energy against its maximum size are represented in Fig.\,\ref{time and energy}.

\begin{figure}[h]
\centering
\includegraphics[width=0.95\linewidth]{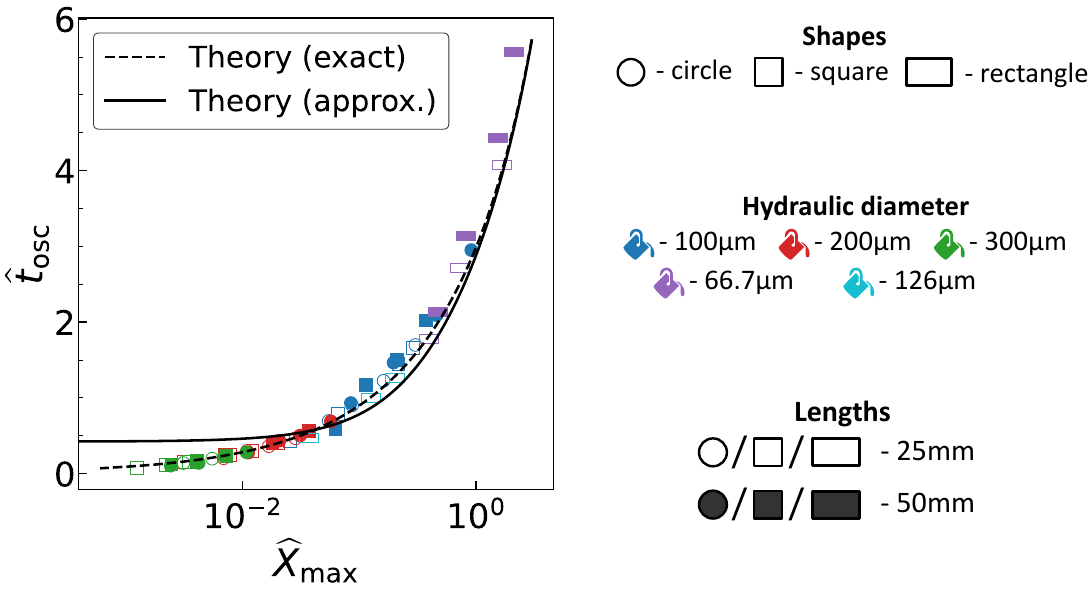}
\caption{The approximate solution to Eq.\,\ref{eqn t_osc} employing Eqs.\,\ref{eqn_approx1}\,and\,\ref{eqn_approx2} is represented using the continuous line, while the exact solution to Eq.\,\ref{eqn t_osc} is represented using dashed line. The approximate solution is with $n$ upto 3 in Eq.\,\ref{eqn_approx1}. The markers are the experiments performed.}
\label{figure t_star_approx}
\end{figure}

While the exact solution to the Lambert $W$ function in Eq.\,3 of the main manuscript accurately captures the experiments, below we provide the approximation to help solve the equation with regular/popular functions. The following are the approximate relations:
\begin{gather}
\tu{W}_0 (\psi) = \sum_{n=1}^\infty \frac{(-n)^{n-1} \psi^n}{n!},\,\, |\psi| \leq \frac{1}{e} \,\,\tu{and} \label{eqn_approx1} \\
\tu{W}_{-1} (\psi) \approx \log(-\psi) - \log(-\log(-\psi)) + \log(-\log(-\psi))/\log(-\psi)...
\label{eqn_approx2}
\end{gather}
Where $\psi=-e^{-(1+\widehat{X}_\tu{max})}$. Figure\,\ref{figure t_star_approx} shows the solution to the above equations for $n$ upto 3 with
\begin{gather}
    \widehat{t}_\tu{osc} = \tu{W}_0 (\psi) - \tu{W}_{-1} (\psi). 
    \label{eqn t_osc}
\end{gather}

\section{Laser energy absorbed by the liquid and threshold energy for bubble formation}
The channel's geometry, such as, wall thickness, material, cross-section shape and dimension can influence the laser energy available for the liquid to absorb. For energy calibration, the energy transmitted through the channel ($E_\tu{trans}$) was measured. We define an energy ratio, $\epsilon_\tu{geom}$, estimated by measuring the energy difference without the channels and channels filled with water. Therefore, if $E_\tu{supp}$ is the energy supplied to the channel, then for channels filled with water, $E_\tu{trans} = \epsilon_\tu{geom} E_\tu{supp}$. The channels filled with water was used as reference to measure energy absorbed in experiments since water is transparent to 532\,nm wavelength and matches the refractive index of the aqueous dye (referred to as liquid). 

For experiments with the aqueous dye, a portion or a whole the energy, $\epsilon_\tu{geom} E_\tu{supp}$, will be absorbed based on the absorption coefficient of the liquid and the distance the light will travel through the liquid. Therefore the energy transmitted through the channel with the liquid is, $E_\tu{trans} = \epsilon_\tu{L} \epsilon_\tu{geom} E_\tu{supp}$, where $\epsilon_\tu{L}$ is the transmission ratio of the liquid. Therefore using the Beer–Lambert law \cite{beerlambertlaw}, the absorbance ($A_\tu{abs}$) of the liquid can be calculated as, 
\begin{equation}
    A_\tu{abs} = -\log_{10}(\epsilon_\tu{L}) = -\log_{10}(E_\tu{trans} / (\epsilon_\tu{geom} E_\tu{supp})).
    \label{Aabs}
\end{equation}

\begin{figure}[h]
\centering
\includegraphics[width=0.95\textwidth]{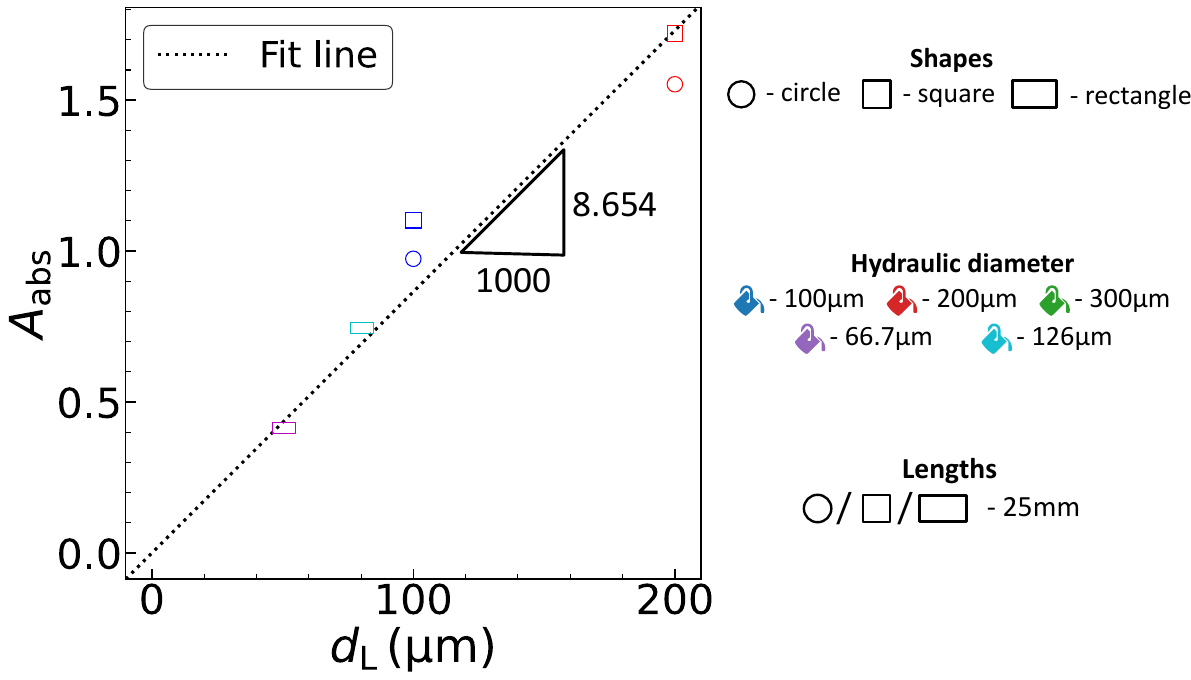}
\caption{The absorbance of the liquid ($A_\tu{abs}$) against the distance the light travels through the liquid ($d_\tu{L}$). The slope represents the absorption coefficient. The marker shapes represent the shape of the cross-section and the colors represent the dimension.}
\label{AabsandE}
\end{figure}

Furthermore, the absorbance can be written as, $A_\tu{abs} = \alpha \,d_\tu{L}$, where $\alpha$ is the absorption coefficient and $d_\tu{L}$ is the distance the light will travel through the liquid. $d_\tu{L}$ is equal to the hydraulic diameter ($d_\tu{h}$) for circular and square channels. For rectangles it is the shorter edge length of the cross-section. Since the red dye (RD81, Sigma-Aldrich) concentration in the liquid is consistent through this work (0.5 wt\%), therefore $\alpha$ is a constant. The $\alpha$ of the liquid is estimated experimentally by measuring the transmitted light as, $\alpha = A_\tu{abs}/d_\tu{L}$. Figure \ref{AabsandE} shows the estimation of $\alpha$ from experiments using the $A_\tu{abs}$ against $d_\tu{L}$ line slope.

For $r\in[-d_\tu{L}/2, d_\tu{L}/2]$ is the radial position from the channel's geometric centre, the laser energy absorbed at the laser focal spot is required to estimate the threshold energy for bubble formation. An estimation for the transmitted energy as a function of $r$ can be represented as, $E_\tu{trans} = 10^{-\alpha (d_\tu{L}/2+r)} \epsilon_\tu{geom} E_\tu{supp}$. By further differentiating it, we get,
\begin{equation}
    \tu{d}E_\tu{trans} = -\epsilon_\tu{geom} E_\tu{supp} \alpha\log(10) 10^{-\alpha (d_\tu{L}/2+r)} \,\tu{d}r.
\end{equation}
In the above expression for $r>d_\tu{L}/2$ is where the energy meter is positioned to calibrate the transmitted energy. Thus $r=0$ correspond to the position of the laser spot, and $\tu{d}E_\tu{trans} (r=0)$ is the energy absorbed at the laser spot. This absorbed energy will be used for sensible and latent heat of the liquid, resulting in the energy balance, $\tu{d}E_\tu{trans} (r=0) = -\pi w_0^2 \rho_\tu{L} (c_p\Delta T + H_\tu{L}) \,\tu{d}r$. Where $2w_0$ is the laser spot diameter, $c_p$ is the specific heat, $\Delta T$ is the rise in temperature and $H_\tu{L}$ the latent heat of vaporization. Thus the energy at the laser spot can be expressed as,
\begin{equation}
    \epsilon_\tu{geom} E_\tu{supp} = \frac{\pi w_0^2 \rho_\tu{L} (c_p\Delta T + H_\tu{L})} {\alpha\log(10) 10^{-\alpha (d_\tu{L}/2)}}.
\end{equation}
In other words, the total energy absorbed by the liquid, $E_\tu{abs}$, can be theorized as follows,
\begin{equation}
    E_\tu{abs} = (1-\epsilon_\tu{L})\epsilon_\tu{geom} E_\tu{supp} = (1-10^{-\alpha d_\tu{L}})\frac{\pi w_0^2 \rho_\tu{L} (c_p\Delta T + H_\tu{L})} {\alpha\log(10) 10^{-\alpha (d_\tu{L}/2)}}.
    \label{Eabs}
\end{equation}

In the above theoretical expression for the absorbed laser energy, by substituting the values of properties of the liquid we can theoretically calculate the threshold energy for bubble formation ($E_\tu{th}$). For a bubble to form, we use the spinodal temperature of water as the necessary condition - temperature at which the water explosively turns into vapor. Thus as the water reaches the spinodal temperature (577\,K) \cite{10.1007/s10953-009-9417-0}, $E_\tu{th} = E_\tu{abs}$. Since the laser pulse width is 4\,ns, the energy from the laser is transferred to the liquid within a short amount of time, which in turn vaporizes the liquid.

\section{Analytical solution to oscillating flow with sinusoidal pressure gradient} \label{analytical solution}
The equations discussed in this section are adapted from Wang \cite{10.1007/s11242-016-0652-8}.\\\\
\textbf{Circular channels} The velocity profiles within a channel with circular cross-section is as follows,
\begin{equation}
    U(r,t) = \frac{(\Delta{p}/\Delta{x}) d_\tu{h}^2}{4\mu_e \sigma^2} [1 + c_1 I_0(2\sigma r/d_\tu{h}) + c_2 K_0(2\sigma r/d_\tu{h})]\, \exp(i\omega t),
\end{equation}
where, $\sigma = \sqrt{k+i\widehat{\omega}}$ and
\begin{equation}
    c_1 = \frac{K_0(-\sigma)-K_0(\sigma)}{I_0(-\sigma)K_0(\sigma) - I_0(\sigma)K_0(-\sigma)},\,
    c_2 = \frac{I_0(\sigma)-I_0(-\sigma)}{I_0(-\sigma)K_0(\sigma) - I_0(\sigma)K_0(-\sigma)}.
\end{equation}
The $I_n$ and $K_n$ are the modified Bessel functions, and $\Delta p / \Delta x = p_\infty/l_\tu{L}$ is the pressure gradient along channel axis. Thus the flow is governed by two parameters, a non-dimensional frequency, $\widehat{\omega}=\rho_\tu{L}d_\tu{h}^2\omega/(4\phi\mu_e)$, and a porous medium factor, $k=\mu/(\mu_e D)$. $D$ is the Darcy number, $\mu_e$ is the effective viscosity and $\phi$ the channel porosity. For pure fluid flows, $D \to \infty$, $\mu=\mu_e$ and $\phi=1$ \cite{10.1063/1.2792323}. Thus the transient mean flow velocity is,
\begin{equation}
    U(t) = \frac{4}{\pi d_\tu{h}^2} \int_0^{d_\tu{h}/2} U(r,t)\, 2\pi r\tu{d}r  = \frac{2\,(\Delta p/\Delta x)}{\mu_e \sigma^2}\, \mathbb{Re}\Bigl\{[0.5 + c_1I_1(\sigma)/\sigma]\, \exp(i\omega t)\Bigr\},
\end{equation}
where $\mathbb{Re}$ is the real part of a complex number.

\begin{figure}[h]
\centering
\includegraphics[width=0.95\textwidth]{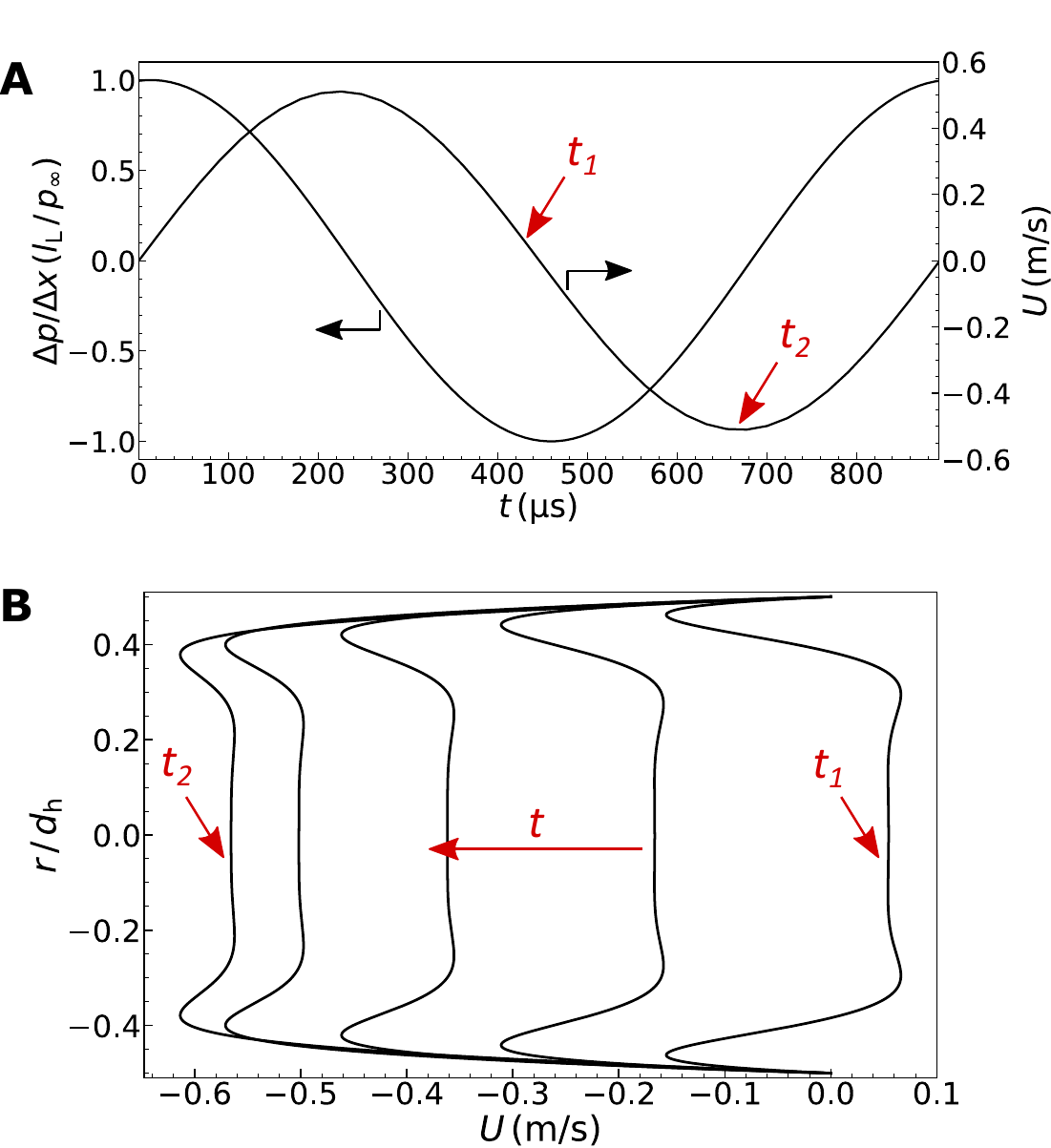}
\caption{(A) This representative analytical solution is for $d_\tu{h}=300\,\upmu$m, $L=50\,\tu{mm}$ and $t_\tu{osc}=892\,\upmu$s. $U$ is the mean flow velocity. (B) The corresponding velocity profiles along the channel's radial direction ($r$) for the solution from (A). $t_1$ and $t_2$ are chosen time window to represent the flow. The direction of $t$ represents the progress in time. The gradient in the velocity at the channel walls suggest a deceleration of the flow following the mean flow reversal (at the start of bubble collapse).}
\label{flow profile}
\end{figure}

Figure \ref{flow profile}A shows the predicted velocity profiles for an oscillating flow with pressure gradient $p_\infty \cos{(\omega t)} /l_\tu{L}$. The representative solution is for $d_\tu{h}=300\,\upmu$m, $L=50\,\tu{mm}$ and $t_\tu{osc} = 892\,\upmu$s. Figure \ref{flow profile}B illustrates the velocity profiles along the channel's radial direction.

\clearpage

\clearpage
\bibliographystyle{apsrev4-2}
\bibliography{references}